%% file: main.tex
\documentclass[conference]{IEEEtran}

  	\usepackage[pdftex]{graphicx}
  	\graphicspath{{../pdf/}{../jpeg/}}
	\DeclareGraphicsExtensions{.pdf,.jpeg,.png}

	\usepackage[cmex10]{amsmath}
    \usepackage{subcaption}
    \usepackage[hyperfootnotes=false]{hyperref}
    \usepackage{amsmath}
    \usepackage{amsfonts}
    \usepackage{hyperref}
    \usepackage{xcolor}
	\usepackage{algorithmic}
	\usepackage{array}
	\usepackage{mdwmath}
	\usepackage{mdwtab}
	\usepackage{eqparbox}
    \usepackage{array}
    \usepackage{multirow}
    \usepackage{makecell}
    \usepackage{nicematrix}
     \usepackage{booktabs} 
 
	\usepackage{url}
\begin{document}

\title{\LARGE Z-upscaling: Optical Flow Guided Frame Interpolation for Isotropic Reconstruction of 3D EM Volumes}


 \author{\authorblockN{Fisseha A. Ferede\authorrefmark{2}\authorrefmark{1}, Ali Khalighifar\authorrefmark{2}, Jaison John\authorrefmark{2}, Krishnan Venkataraman\authorrefmark{2}, Khaled Khairy\authorrefmark{2}}
 \authorblockA{\authorrefmark{2}Center for Bioimage Informatics, St. Jude Children’s Research Hospital, Memphis, TN 38105, USA
}
 \authorblockA{\authorrefmark{1}Department of Electrical and Computer Engineering, University of Memphis, Memphis, TN 38152, USA\\*Corresponding authors: fissehaad@gmail.com, khaled.khairy@stjude.org}}

\maketitle
\footnotetext{This work has been submitted to the IEEE for possible publication. Copyright may be transferred without notice, after which this version may no longer be accessible.}

\begin{abstract}
We propose a novel optical flow based approach to enhance the axial resolution of anisotropic 3D EM volumes to achieve isotropic 3D reconstruction. Assuming spatial continuity of 3D biological structures in well aligned EM volumes, we reasoned that optical flow estimation techniques, often applied for temporal resolution enhancement in videos, can be utilized. Pixel-level motion is estimated between neighboring 2D slices along z, using spatial gradient flow estimates to interpolate and generate new 2D slices resulting in isotropic voxels. We leverage recent state-of-the-art learning methods for video frame interpolation and transfer learning techniques, and demonstrate the success of our approach on publicly available ultrastructure EM volumes.
\end{abstract}


\IEEEoverridecommandlockouts

\IEEEpeerreviewmaketitle


\section{Introduction}

3D electron microscopy is a powerful technique for visualizing and analyzing the three-dimensional structure of biological samples at nanometer scale. However, the axial resolution (the ability to distinguish between two points along the optical or z-axis) is often lower than the lateral resolution (the ability to distinguish points in the plane perpendicular to the optical axis or $x-y$ plane) due to inherent limitations of typical EM imaging systems. This results in anisotropic voxel dimensions. Resolution in x y is typically sufficient to identify ultrastructure in EM images.
 

Achieving isotropic, or close to isotropic, resolution is desirable to extend the quantitative value of EM in the third dimension \cite{plaza2014toward}. Even  EM for modalities that produce images at sufficiently high resolution in any dimension, image acquisition time can be prohibitive (years for larger volumes). This further motivates a process in which EM acquisitions are performed at low resolution, and that can take advantage of a computational post-acquisition upscaling step. 

Several deep-learning based image processing techniques have been proposed in recent years for 3D isotroptic reconstruction of EM volumes. Heinrich et al. \cite{heinrich2017deep} utilized 3D Convolutional Neural Network (3D CNN) and U-Net architectures for isotropic super-resolution reconstruction of EM volumes. However, its training process depends on the availability of isotropic ground truth EM data, which can be challenging to obtain. Lee et al.\cite{lee2023reference} applied 2D diffusion models for isotropic reconstruction of 3D EM volumes by approximating degradations along the $z-axis$ with 2D degradations of neighboring ZY (or ZX) slices. While this process doesn't rely on isotropic ground-truth or point spread function (PSF) of the imaging system for training, assuming identical degradation functions for lateral and axial dimensions may fail due to differences in aberration effects across these axes that can contribute to different degradation functions \cite{hovhannisyan2008characterization}. 

In this work, we introduce a novel, learning-based spatial interpolation technique for isotropic 3D reconstruction of EM volumes. Unlike conventional methods, our approach eliminates the dependence on isotropic ground truth data or the PSF of the imaging system, making it versatile across various imaging modalities. Additionally, it can be effectively trained on video datasets, overcoming limitations of scarce medical data for training. Through extensive experimentation, we demonstrate that our method achieves superior qualitative and quantitative isotropic reconstruction surpassing other interpolation based techniques. 

\input{Figure_docs/vizual_eg_result}

\input{Figure_docs/architecture}

The main contributions of this paper can be summarized as follows:
 \begin{itemize}


    \item  We formalized the isotropic reconstruction of 3D medical image volumes across different imaging modalities as a flow-based spatial interpolation problem, overcoming the need for isotropic ground-truth images and making our method adaptable across various imaging modalities.

    \item We adopt the architecture of FILM \cite{reda2022film}, a video interpolation technique, and transfer learning techniques for isotropic reconstruction of 3D EM volumes.

    \item We experimentally demonstrate that large-scale video datasets, typically used for temporal resolution enhancement, can effectively train robust spatial interpolation models for medical images, overcoming the limited availability of medical image datasets.


\end{itemize}

\section{Methodology}

\subsection{Temporal Interpolation}

For a given video frame sequence, $\nu \in R^{w\times h\times t}$, where $w\times h$ represent the spatial dimensions and $t \in {1,2,..,T}$ the temporal dimension consisting of discrete image sequences $I_1,I_2,..,I_T$, let $E(x,y,t)$ represent the dense and pixel-level brightness values of the sequence in space-time domain. The partial derivative of this intensity function, $E(x,y,t)$, with respect to time is given as:

        \begin{align}\label{eq:01}    
          \frac{\partial E(x,y,t)}{\partial t} = \frac{\partial E}{\partial x} \frac{ dx}{dt} + \frac{\partial E}{\partial y} \frac{ dy}{dt} +  \frac{dE}{dt}
        \end{align}

Such that the terms $\frac{ dx}{dt}$ and $\frac{ dy}{dt}$ in Eq. \ref{eq:01} represent the relative movements of temporally neighboring image frames in the $x$ and $y$ directions, respectively. These terms are commonly known as optical flow. Several learning based methods have been introduced to estimate 2D and dense optical flows from sequences of video frames \cite{dosovitskiy2015flownet, teed2020raft, ferede2023sstm}.

Inspired by optical flow estimation techniques, flow guided video interpolation techniques \cite{reda2022film, kalluri2023flavr} aim at enhancing the temporal resolution of video frames by synthesizing non-existing video frames. Given temporally neighbouring video frames $I_{0}$ and $I_{1}$, such methods first estimate the bi-directional optical flow fields, $F_{{I_t}\rightarrow I_{0}}$ and $F_{{I_t}\rightarrow {I_1}}$, between an unknown frame $I_t, t \in (0,1)$ and the input frames. These bi-directional flow estimates are then warped onto the input images using a backward warping function $\phi(\tau(I_0), F_{{I_t}\rightarrow I_{0}})$ where $\tau(\cdot)$ represents a feature extraction function applied on the input images. A decoder architecture then learns to synthesize the unknown in-between frame, $I_t$, from these warped estimations.

\subsection{Spatial Interpolation for 3D Isotropic Reconstruction}

The fundamental assumption in flow based video interpolation techniques is that there is a pixel level continuity / correspondence between temporally neighbouring image frames such that most pixels and structural elements in the reference frame, $I_0$, have corresponding representations in the target frame, $I_1$, that can be represented by a 2D flow field. 

In 3D medical image volumes, 2D slices of a given 3D volume, across a given axis, represent cross-sectional views of 3D biological structures. These cross-sections depict  boundaries of smooth and continuous surfaces that make up such 3D structures. For example, the 2D slices of a simple spherical structure in 3D space would appear as a sequence of concentric circles which can be represented by 2D spatial flows. Thus, the above formulation in Eq. \ref{eq:01} should hold true for spatially neighbouring 2D slices of $Z_1$ and $Z_2$ of a given 3D volume in space, $E(x,y,z)$, such that $\frac{ dx}{dz}$ and $\frac{ dy}{dz}$ represent the spatial flows across the axial axis of spatially neighbouring $z$ slices.  

\input{Figure_docs/vizual_results2}

\input{Tables/schedule}
\input{Tables/results}

\subsection{FILM for 3D Isotropic Reconstruction}
We leverage FILM \cite{reda2022film} architecture, a recent state-of-the-art learning-based method for temporal interpolation task in videos, for 3D isotropic reconstruction of 3D EM volumes. We further applied transfer learning techniques which included fine-tuning pretrained (on video) models on EM datasets and training from scratch only on EM datasets. 

Fig \ref{fig:architecture} shows the overall architecture of our implementation of FILM for 3D isotropic reconstruction. First, we extract pyramid features of the coarse-to-fine representation of two neighbouring $z$ slice inputs, $Z_1$ and $Z_3$. This is done by down-scaling the original $xy$ resolutions of each slice by different levels (the finest representation being the original full resolution). Features at different resolution level are then concatenated to form pyramid features that can collectively capture small and large spatial flows. Second, extracted pyramid features are used to estimate bidirectional spatial flows, $F_{Z_2 \rightarrow Z_1}$ $F_{Z_2 \rightarrow Z_3}$, from the unknown frame $Z_2$ to the input images. Third, these spatial flows are warped onto the feature representations of input images using a warping function $\phi(\cdot)$ at different resolution levels. Lastly, the warped features are fused together and passed through a U-net based decoder architecture to reconstruct the unknown frame $Z_2$. By recursively applying this interpolation on newly interpolated frames, we can achieve up-sampling at $2^n$ order.
\section{Experiments}
\subsection{Dataset}
In this experiment, we used several image datasets across different modalities for training as well as testing of 3D isotropic reconstruction of medical image volumes. 

\subsubsection*{\textbf{Vimeo-90k}}
Vimeo-90k \cite{xue2019video} for interpolation is a standard benchmark dataset curated for the task of middle frame video frame interpolation. The dataset consists of 73,171 3-frame sequences constructed from 15k video clips with diverse ranges of motion and scene variations. In this report, we used the training set of Vimeo-90k which consists of 51,314 sequences with resolution of 448$\times$256 for training.

\subsubsection*{\textbf{FlyEM}} 
FlyEM Hemibrain \cite{scheffer2020connectome} is a large Electron Microscope dataset focused on mapping a detailed connectome of a substantial portion of the \textit{Drosophila melanogaster} central brain. This dataset has an isotropic voxel size of $8\times8\times8 nm$. We cropped 120 non-overlapping sub-volumes from the original dataset with size of 2001$\times$256$\times$448 voxels in the $z,$ $x$ and $y$ directions, respectively. From each sub-volume, we generated 1000 sequences of 3 spatially neighbouring frames. Specifically, each sub-volume $V_i | i\in \{1,2,..,120\}$ is used to generate a sequence $S_i = \{ \{I_{z_1}, I_{z_2}, I_{z_3}\}, \{I_{z_3}, I_{z_4}, I_{z_5}\}, ... , \{I_{z_{1999}}, I_{z_{2000}}, I_{z_{2001}}\} \}$.
 
\subsubsection*{\textbf{FIB-25 EM}}
FIB-25 \cite{takemura2015synaptic} is a large EM dataset of the \textit{Drosophila melanogaster} medula generated using \textit{Focused Ion Beam Scanning Electron Microscopy (FIB-SEM)}. The dataset has isotropic $8\times8\times8 nm$ voxel size  covering a volume of a $52 \times 53 \times 65 \mu m$. We extracted a 256 sequences of spatially neighbouring z-slices with lateral dimension of $256\times256$ voxels from the full dataset, which contains numerous intricate features. This dataset is used for testing trained models only. 

\subsection{Training Schedule}
Training and fine-tuning FILM \cite{reda2022film} for isotropic reconstruction of medical image volumes was done on 4 Tesla V100 GPUs. Our training schedule includes training from scratch and fine-tuning stages on EM and video datasets to build three different methods namely, FILM \cite{reda2022film} (trained on video), FILM-sc-EM (trained on EM) and FILM-ft-EM (trained on video and fine-tuned on EM). In all of these training stages, we used a batch size of 8, style loss function, seven pyramid levels (finest and coarsest resolutions being $256\times448$ and $4\times 7$, respectively) as described in FILM \cite{reda2022film}. Moreover, data augmentation is applied on the training data using random rotations, flips, and crops. Table \ref{tab:table-schedule} shows the summary of these training stages.

\subsection{Evaluation}
We used standard evaluation metrics for image reconstruction quality measurement namely, Peak Signal-to-Noise Ratio (PSNR) and Structural Similarity Index (SSIM). We evaluated our approach by skipping specific frames (used as ground truth) and measuring the quality of our reconstruction for these skipped frames. For example, for $\times4$ up-sampling factor, given $Z1$ and $Z_5$, we estimate $\hat{Z}_2, \hat{Z}_3$ and $\hat{Z}_4$ and measure the reconstruction qualities by comparing them with skipped frames $Z_2, Z_3$ and $Z_4$.


\section{Discussion}
In Table \ref{tab:table-resul}, we reported the PSNR and SSIM results of our methods, FILM-sc-EM and FILM-ft-EM, by comparing against bicubic interpolation and other learning based interpolation methods \cite{reda2022film, bao2019depth, wang2022stdin}. Our methods achieved overall superior performance on the FIB-25 and Cremi test datasets. Moreover, in Fig \ref{fig:main} and Fig \ref{fig:results}, we demonstrated the visual qualities of our reconstructed 2D slices by comparing against the ground truth. Our approach is highly scalable, making it well-suited for deployment in large-scale bioimage processing environments and adaptable to real-world applications.

Future research directions include learning spatial interpolation across multiple z-slices for globally coherent and more accurate reconstructions, as well as achieving arbitrary up-sampling factors beyond the limitation of $2^n$ order. Additionally, this technique could also be extended to other imaging modalities, such as confocal and light sheet microscopy. Software code, example data volumes and instructions:  \href{https://github.com/Fisseha21/Z-upscaling}{https://github.com/Fisseha21/Z-upscaling}.

\section{Conclusion}
We presented a novel learning-based spatial interpolation technique for the 3D reconstruction of isotropic EM volumes, offering a robust and versatile solution that operates without the need for isotropic ground truth data or any information about the degradation function of the imaging system. By leveraging video datasets along with EM datasets for training, our approach effectively overcomes the limitations of scarce biomedical data, broadening its applicability to a wide range of imaging modalities while achieving robust and highly accurate reconstruction. 

\section{Conflict of Interest}
This research was supported by the American Lebanese Syrian Associated Charities (ALSAC) of St. Jude Children's Research Hospital. The authors declare no conflict of interest.

\bibliographystyle{IEEEtran}
\bibliography{biblio_traps_dynamics}

\end{document}

%% file: Figure_docs/vizual_eg_result.tex
\begin{figure}[ht]
    \centering
    \begin{subfigure}{0.22\textwidth}
        \centering

        \includegraphics[ width=1\textwidth]{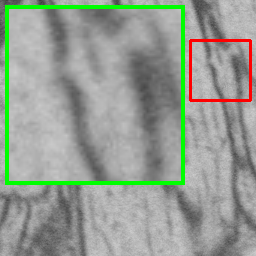}
        \caption{Ground truth}
        \label{fig:sub-a}
    \end{subfigure}
    \hfill
    \begin{subfigure}{0.22\textwidth}
        \centering
        \includegraphics[width=\textwidth]{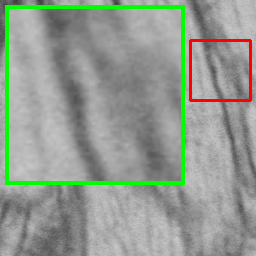}
        \caption{Bicubic}
        \label{fig:sub-b}
    \end{subfigure}
    
    \medskip
    
    \begin{subfigure}{0.22\textwidth}
        \centering
        \includegraphics[ width=\textwidth]{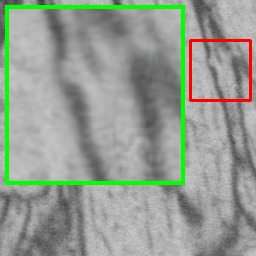}
        \caption{Trained on Video dataset}
        \label{fig:sub-c}
    \end{subfigure}
    \hfill
    \begin{subfigure}{0.22\textwidth}
        \centering
        \includegraphics[width=\textwidth]{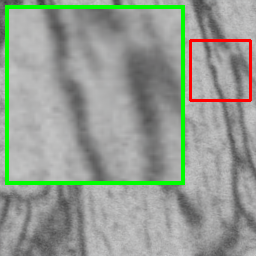}
        \caption{Trained on EM dataset}
        \label{fig:sub-d}
    \end{subfigure}
    \caption{Sample reconstruction of FIB-25 EM data for $\times8$ up sampling. Reconstruction within the red box is zoomed in and overlaid on the green box region for clearer comparison.}
    \label{fig:main}
\end{figure}

%% file: Figure_docs/architecture.tex
\begin{figure*}[ht]
    \centering
    \includegraphics[width=0.9\textwidth]{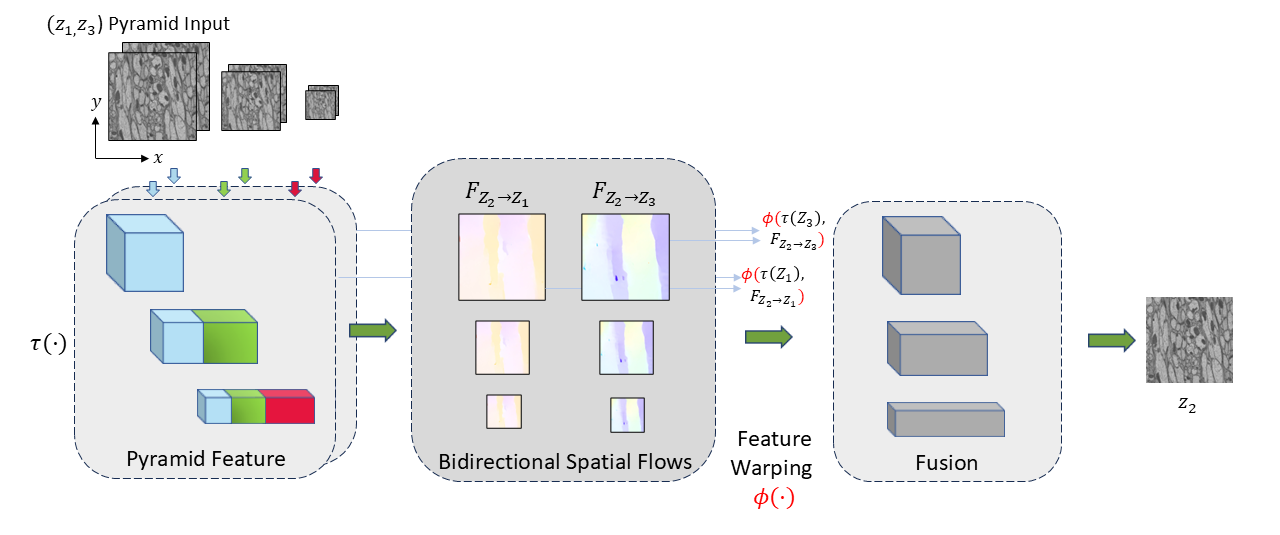}  
    \caption{Architecture of FILM for 3D isotropic reconstruction. Feature extractor,$\tau (\cdot)$, takes multi-resolution pyramid input of a pair of neighbouring $Z$ slices, $(Z_1, Z_3)$ to extract pyramid features. These pyramid features are then used to compute bi-directional spatial flows from unknown frame $Z_2$ to the input slices. Estimated spatial flows are warped onto the input frames and decoded using U-net decoder to reconstruct middle frame output, $Z_2$. }
    \label{fig:architecture}
\end{figure*}

%% file: Figure_docs/vizual_results2.tex
\begin{figure*}[ht]
    \centering
    \begin{subfigure}{0.49\textwidth}
        \centering
        \includegraphics[width=\textwidth]{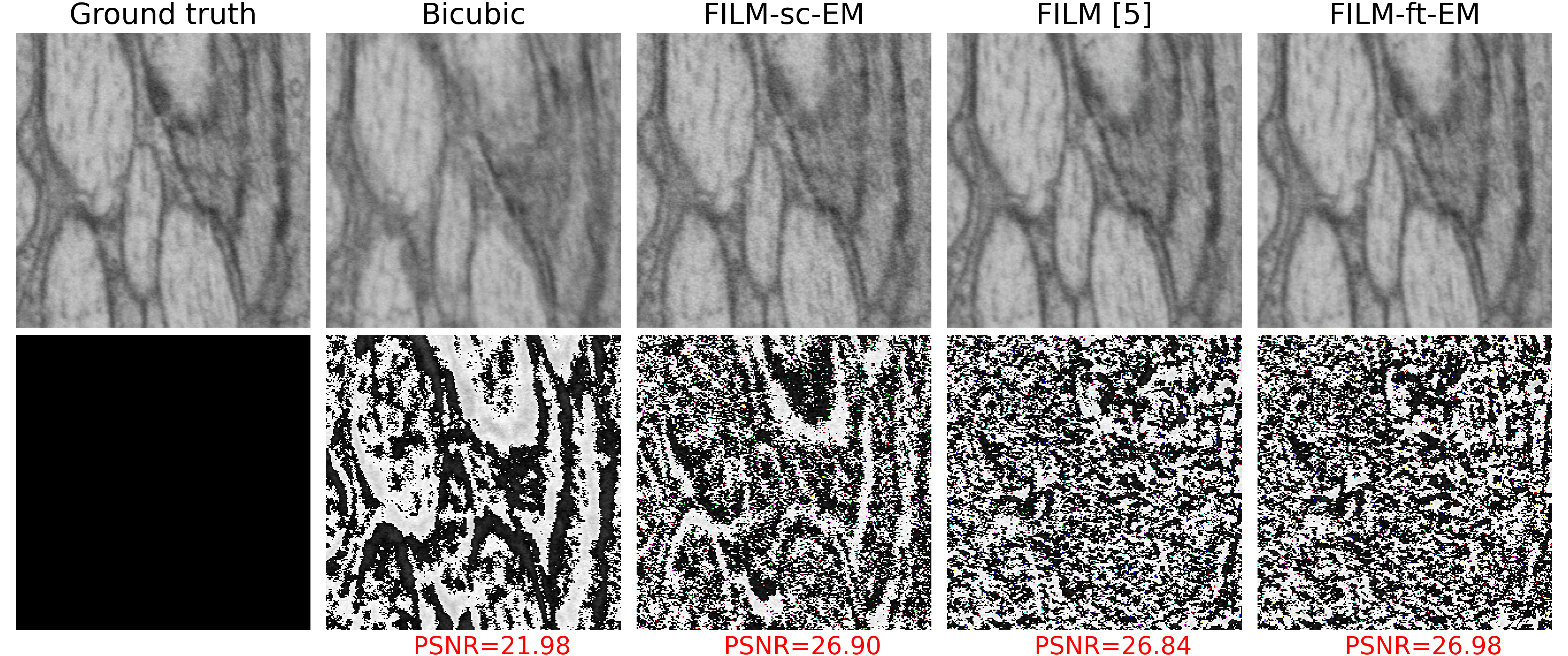}
        \label{fig:resu-a}
    \end{subfigure}
    \hfill
    \begin{subfigure}{0.49\textwidth}
        \centering
        \includegraphics[width=\textwidth]{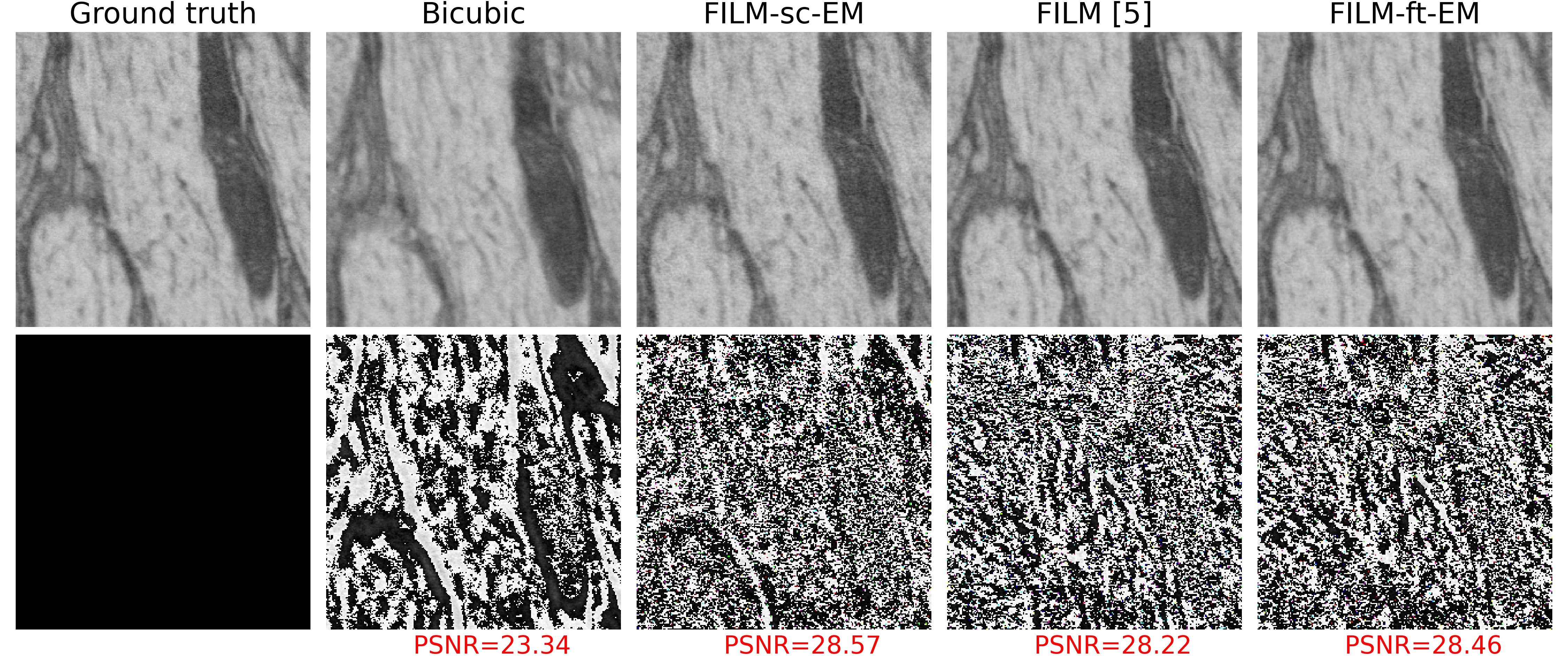}
        \label{fig:resu-b}
    \end{subfigure}
    \medskip
    \caption{Sample visualization results on FIB-25 dataset. The first rows demonstrate the reconstructed z slices along with the corresponding ground truths. The second row shows pixel-wise $L1$ norm error map and PSNR values of the reconstructed frames. White regions in the error map signify erroneous regions.}
    \label{fig:results}
\end{figure*}

%% file: Tables/schedule.tex
\begin{table*}[ht]
    \caption{Summary of training and fine-tuning schedule. Training datasets $\mathbb{V}$ and $\mathbb{F}$ refer to Vimeo90k and FlyEM training datasets, respectively. $\mathbb{V+F}$ refers to training on Vimeo90k video dataset followed by fine-tuning on FlyEM.}
   \begin{tabular*}{\textwidth}{@{\extracolsep{\fill}} ll *{8}{c} }
   \cmidrule{1-5} 
   \multirow{2}{*}{\kern21pt Method }
   &\multirow{1.5}{*}{ \kern11pt Training}
   &\multirow{1.5}{*}{Imaging}
   &\multirow{1.5}{*}{Size}
   &\multirow{1.5}{*}{ Number of}\\

   \multirow{2}{*}{\kern10pt }
   &\multirow{1.5}{*}{ \kern14pt dataset}
   &\multirow{1.5}{*}{ }
   &\multirow{1.5}{*}{}
   &\multirow{1.5}{*}{iterations}\\
   
   \cmidrule{1-5}

   \kern15pt FILM \cite{reda2022film}  & \kern28pt $\mathbb{V}$ & Video & [448, 256] & 3M \\ 
   
   \kern15pt FILM-sc-EM & \kern28pt $\mathbb{F}$ & EM & [448, 256] & 3M\\
   
   \kern15pt FILM-ft-EM & \kern20pt $\mathbb{V+F}$ & Video + EM & [448, 256]  & 1.5M + 1.5M \\
    
   \cmidrule{1-5}
   \end{tabular*} 
   \label{tab:table-schedule} 
\end{table*}

%% file: Tables/results.tex
\begin{table*}[ht]
    \caption{Performance of flow based interpolation methods for 3D reconstruction of EM volumes. Methods are evaluated on FIB-25 and Cremi-C ($\mathbb{C}$) test datasets. Cremi-C dataset is used for both training and testing in case of STDIN \cite{wang2022stdin}. PSNR and SSIM performance measures are reported. Best results are highlighted in \textbf{bold}.}
    \begin{tabular*}{\textwidth}{@{\extracolsep{\fill}} ll *{8}{c} }
        \cmidrule{1-9}
        \multirow{2}{*}{\kern18pt Up-sampling}
         &\multirow{2}{*}{\kern5pt Method}
         &\multirow{2}{*}{\kern6pt Training\kern1pt}
         &\multicolumn{2}{c}{ FIB-25 EM (test)} 
           & \multicolumn{2}{c}{Cremi EM (test)} \\
         \cmidrule{4-5} \cmidrule(lr){6-7} \cmidrule(lr){8-9}

          \kern30pt factor & & \kern5pt data & PSNR $(\uparrow)$ & SSIM $(\uparrow)$ & PSNR $(\uparrow)$ & SSIM $(\uparrow)$ \\
        \cmidrule{1-9}

    & Bicubic & \kern7pt - &  30.42  & 0.80 & 18.29 & 0.29 \\
    & DAIN \cite{bao2019depth} & \kern7pt $\mathbb{V}$&  - & - & 15.24 & 0.32 \\
    
    & FILM \cite{reda2022film} & \kern8pt $\mathbb{V}$  &  31.12 & 0.80  & 19.73 & 0.40  \\
    
    & STDIN \cite{wang2022stdin} & \kern7pt $\mathbb{C}$ &  - & - & 17.83 &  \textbf{0.43}  \\
    
   \kern30pt $\times$2 & FILM-sc-EM & \kern7pt $\mathbb{F}$ & \textbf{31.52} & \textbf{0.82 }& 17.21&  0.30 \\
   
   & FILM-ft-EM &\kern5pt $\mathbb{V+F}$ & 31.16  & 0.81 & \textbf{19.75} & 0.40 \\ 
\cmidrule{1-9}
        & Bicubic & \kern7pt - & 26.46 & 0.68 & 17.41 & 0.25\\
        & FILM \cite{reda2022film} & \kern7pt $\mathbb{V}$ & 28.15  & 0.68  & \textbf{18.41} & \textbf{0.31} \\
        
       \kern30pt $\times$4 & FILM-sc-EM & \kern7pt $\mathbb{F}$ & \textbf{28.83} &\textbf{ 0.74} & 15.69 & 0.25 \\
       
        & FILM-ft-EM & \kern5pt $\mathbb{V+F}$  & 28.31 & 0.69  & 18.40 & \textbf{0.31} \\ 
         
\cmidrule{1-9} 
        & Bicubic & \kern7pt -  & 22.66 & 0.51 & 16.99 & 0.22 \\
        & FILM \cite{reda2022film} & \kern7pt $\mathbb{V}$  & 25.56 &  0.55 & \textbf{17.55} & \textbf{0.26} \\
       \kern30pt $\times$8 & FILM-sc-EM & \kern7pt $\mathbb{F}$& 25.40 & \textbf{0.59} &  14.53 &  0.21\\
       & FILM-ft-EM &\kern5pt $\mathbb{V+F}$ & \textbf{25.74} & 0.57 & 17.53& \textbf{0.26}\\ 
         
\cmidrule{1-9} 
\end{tabular*} 
\label{tab:table-resul}
\end{table*}